\documentclass[aps,prb,twocolumn,amsmath,superscriptaddress]{revtex4-2}
\usepackage{graphicx}
\usepackage{epstopdf}
\usepackage{xcolor}

\begin{document}

\title{Coupling of the triplet states of a negatively charged exciton in a quantum dot with the spin of a magnetic atom}

\author{L.~Besombes}\email{lucien.besombes@neel.cnrs.fr}
\affiliation{Institut Néel, CNRS, Université Grenoble Alpes and Grenoble INP, 38000 Grenoble, France}
\affiliation{Japanese-French lAboratory for Semiconductor physics and Technology (J-FAST)–CNRS–Université Grenoble Alpes–Grenoble
INP–University of Tsukuba, 1-1-1 Tennoudai, Tsukuba, 305-8573, Japan}

\author{S.~Ando}
\affiliation{Institute of Materials Science, University of Tsukuba, 1-1-1 Tennoudai, Tsukuba, 305-8573, Japan}

\author{S.~Kuroda}
\affiliation{Institute of Materials Science, University of Tsukuba, 1-1-1 Tennoudai, Tsukuba, 305-8573, Japan}
\affiliation{Japanese-French lAboratory for Semiconductor physics and Technology (J-FAST)–CNRS–Université Grenoble Alpes–Grenoble
INP–University of Tsukuba, 1-1-1 Tennoudai, Tsukuba, 305-8573, Japan}

\author{H.~Boukari}
\affiliation{Institut Néel, CNRS, Université Grenoble Alpes and Grenoble INP, 38000 Grenoble, France}

\date{\today}

\begin{abstract}

Two electrons in a quantum dot (QD) can form triplet states. We analyze the exchange coupling of the triplet states of the negatively charged exciton in a QD (X$^-$, two electrons and one hole) with the spin of a magnetic atom (Mn). Two techniques are used to access the spin structure of this magnetic complex: the resonant excitation of the excited states of X$^-$-Mn and the analysis of the emission of a negatively charged biexciton in a magnetic dot (XX$^-$-Mn). The photo-luminescence (PL) excitation of X$^-$-Mn reveals excited states with a circularly polarized fine structure which strongly depends on the Mn spin state S$_z$ and gives rise to negative circular polarization emission. This fine structure arises from the coupling of the triplet states of an excited charged exciton with the Mn (X$^{-*}$-Mn) and its S$_z$ dependence can be described by a spin effective model. The recombination of XX$^-$-Mn leaves in the dot a charged exciton in its excited state and the PL structure is controlled by the coupling of triplet states of X$^{-*}$ with the Mn spin. An analysis of the polarization and magneto-optic properties of this emission gives access to the electron-hole exchange interaction within the triplets states. Comparing the fine structure of the singlet X$^-$-Mn and of the triplets of X$^{-*}$-Mn we can independently study the different source of anisotropy in the QD: the valence band mixing and the exchange interaction in an anisotropic potential. 

\end{abstract}

\maketitle

\section{Introduction}

The spin state of a carrier in a semiconductor quantum dot (QD) or of an individual impurity in a semiconductor host can act as a bit of quantum information. The transfer of quantum information between such solid-state quanta and single photons is extensively studied as it is an essential element towards the establishment of quantum information networks \cite{Lukin2019}. Optically active semiconductor QDs can be key systems for spin-photon quantum coupling \cite{Senellart2022} as in a QD containing a single carrier, the polarization of the optical emission is directly linked to the spin of the resident carrier. 

Confined carriers in QDs can be exchanged coupled with embedded magnetic elements. This exchange interaction can be exploited to transfer information between the spin of carriers, proposed as efficient qubits in quantum computing devices \cite{Vandersypen2017,Meunier2021}, and a more localized spin on a defect or an impurity. In the case of an optically active QD, this exchange coupling also provides an optical access to a strongly localized spin. This has been demonstrated for some transition metal elements in II-VI and III-V semiconductors \cite{Besombes2004,Kudelski2007,Kobak2014,Smolenski2016,Lafuente2016} and could be extended to other non-optically active individual magnetic defects.

The optical selection rules and the spin dynamics in QDs are strongly influenced by the electron-hole (e-h) exchange interaction. In the three dimensional confinement potential of a dot, the exchange interaction intensifies and usually become anisotropic. This is observed in neutral QDs where an exchange interaction of electron and hole induced by a dot asymmetry splits the exciton radiative doublet and destroys the spin polarization of QD excitons \cite{Flissikowski2001}.  

The e-h exchange interaction is absent for the singlet (S$_0$) of a negatively charged exciton (X$^-$) in the ground state of a QD. However, when at least an electron occupies an excited state of the dot, the two electrons of X$^-$ can also form triplet states (T$_0$,T$_{\pm 1}$). Electron-electron (e-e) and e-h exchange interactions partly remove the spin degeneracy of these triplets. The spin dynamics among the triplets is dominated by the e-h exchange interaction and is at the origin of the observed negative circular polarization in the photoluminescence (PL) of negatively charged QDs \cite{Cortez2002, Ware2005}. The negative circular polarization rate depends on the spin of the resident electron and is an efficient probe of possible optical pumping of the electron spin interacting with the nuclear spin bath of the semiconductor host \cite{Eble2006, Legall2012}.  

The presence of a transition-metal element in the QD significantly modifies the spin structure as both the electrons and the holes are coupled with the localized spin of the magnetic atom through the $sp-d$ exchange interaction \cite{Furdyna1988}. This exchange interaction permits in particular to interface the spin of the atom with single photons. In the case of a charged dot, the simple unpolarized single line emission of X$^-$ transforms into an eleven lines spectra in the presence of single Mn atom \cite{Leger2006}. 

In this work, we analyze how the triplet states of two electrons couple with the spin of a magnetic atom. We investigate the exchange coupling of excited negatively charged excitons (X$^{-*}$, two electrons and one hole) with a Mn atom (S=5/2) in a CdTe/ZnTe QD and analyze how the spin relaxation mechanism of the triplets is affected by the exchange interaction with the Mn spin. PL excitation spectra (PLE) of a negatively charged exciton coupled with a Mn (X$^-$-Mn) exhibits resonances with circularly polarized fine structure arising from a direct excitation of electron spins triplet states exchanged coupled with the Mn spin. The splitting of these absorption resonances depends on the spin state S$_z$ of the magnetic atom. The direct injection of spin polarized e-h pairs on the excited states can give rise to large cross-circularly polarized emission. A model of the circularly polarized fine structure of the PLE is presented.

In some of the charged dots, the optical recombination of the negatively charged biexciton coupled with the Mn (XX$^-$-Mn) is observed. It can be used as a monitor of the optically active triplets states of X$^{-*}$-Mn as they are left behind after the optical recombination of the complex. Its linear polarization and magneto-optical properties are analyzed. The PL of XX$^-$-Mn permits in particular to observe both the isotropic and the anisotropic part of the e-h exchange interaction within the triplets. Different directions of linear polarization are usually observed in the fine structure of the two charged excitonic species, X$^-$-Mn and XX$^-$-Mn. This shows the independence between the principal axis of the strain induced valence band mixing and the shape anisotropy of the dots. The linearly polarized fine structure and the magnetic field dependence of XX$^-$-Mn can be modeled by a spin effective Hamiltonian. A slight reduction of the anisotropic part of the e-h exchange interaction is also observed under a magnetic field applied along the QD growth axis. 

The rest of the paper is organized as follows. After a short presentation of the sample and experiments in section II, we discuss in section III the PL of negatively charged Mn-doped QDs where the emission of charged exciton and charged biexciton can be identified. In section IV, we analyze the circularly polarized fine structure of the resonantly excited triplet states of the negatively charged exciton coupled with the spin of a Mn atom. In section V we discuss the magneto-optics properties of the negatively charged biexciton coupled with a magnetic atom and show how they can be used to access the structure of the excited charged exciton triplet.

\section{Samples and experimental setup}

In this study we use self-assembled CdTe/ZnTe QDs grown by molecular beam epitaxy and doped with Mn atoms. The concentration of Mn was adjusted to obtain QDs containing 0, 1 or a few magnetic atoms. A $n$ doping with aluminum is introduced in the capping layer 30 nm away from the QD layer. When doped with a single magnetic atom these dots permit to optically access the spin of a Mn interacting with negatively charged excitonic complexes. 

Individual QDs were studied by optical micro-spectroscopy at liquid helium temperature (T=4.2 K). The PL was excited with a tunable dye laser, collected by a high numerical aperture microscope objective (NA=0.85), dispersed by a two meters double monochromator and detected by a cooled Si CCD. The sample was inserted in the bore of a vectorial superconducting coil and a magnetic field up to 9 T along the QD growth axis and 2T in the QD plane could be applied. For the PLE measurements, the power of the tunable dye laser was stabilized with an electro-optic variable attenuator.

\section{PL of negatively charged Mn-doped QDs.}

\begin{figure}[hbt]
\centering
\includegraphics[width=1.0\linewidth]{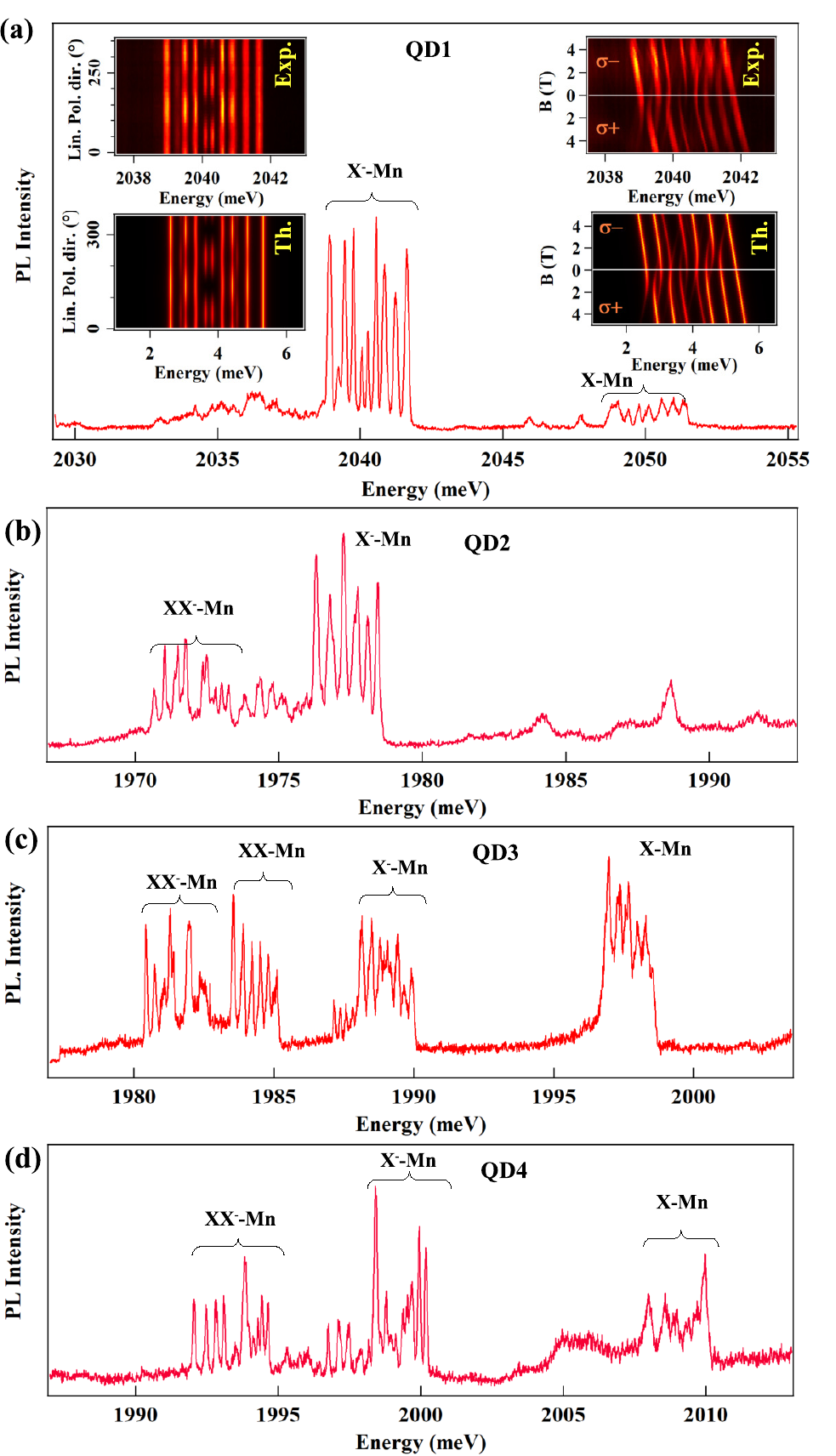}
\caption{PL spectra of QDs observed in a negatively charged Mn-doped sample. X$^-$-Mn can be identified with their linearly polarized fine structure (left insets of (a)) and their low magnetic field dependence (right insets of (a)) of the PL. In many dots, additional PL structures appear on the low energy side of the X$^-$-Mn lines (see XX$^-$-Mn and XX-Mn in QD2 (b), QD3 (c) and QD4 (d)). The parameters used in the model presented in the insets in (a) are $I_{e_1Mn}$=-100 $\mu eV$, $I_{h_1Mn}$=320 $\mu eV$, $\rho_s/\Delta_{lh}$=0.15, $\theta_s=\pi/4$, g$_e$=-0.4, g$_h$=0.5, g$_{Mn}$=2.}
\label{Fig1}
\end{figure}

The typical emission of negatively charged Mn-doped QDs are presented in Fig.~\ref{Fig1}. They present a complex PL spectra which strongly depends on the excitation power. However, among these emission lines, the PL of the ground state of X$^-$-Mn can be identified at zero magnetic field by (i) the number of PL lines and by (ii) the linear polarization dependence of the PL (see for instance QD1 in Fig.~\ref{Fig1}) \cite{Leger2006, Besombes2007, Lafuente2015}.

Let us recall here the origin of the PL structure of X$^-$-Mn. The emitting state in the X$^-$-Mn optical transition has two electrons and one hole coupled to the Mn spin. The effect of the two spin-paired electrons on the Mn can be neglected and X$^-$-Mn is governed by the hole-Mn interaction ${\cal H}_{h-Mn}=I_{h_1Mn}\vec{S}\cdot\vec{J}$. I$_{h_1Mn}$ is the exchange integral of a hole located in the ground state of the dot with the Mn spin. In the heavy-hole approximation the twelve eigenstates of ${\cal H}_{h-Mn}=I_{h_1Mn}S_zJ_z$, $|S_z,J_z\rangle$, are organized as six doublets with defined $S_z$ and $J_z$, with $z$ along the QD growth axis.

In the subspace of the two low-energy heavy-hole states J$_{z}$=$\pm$3/2, a pseudo-spin operator $\tilde{j}$ can be used to take into account a possible valence band mixing (VBM)  \cite{Leger2007,Tiwari2021}. For a VBM induced by an in-plane anisotropy of the strain, the components of $\tilde{j}$ are related to the Pauli matrices $\tau$ by $\tilde{j}_z= \frac{3}{2}\tau_z$ and $\tilde{j}_{\pm}=-2\sqrt{3}e^{-2i\theta_s}\rho_s/\Delta_{lh}\tau_{\pm}$. $\rho_s$ is the coupling energy between heavy-holes and light-holes split by the energy $\Delta_{lh}$ and $\theta_s$ the angle relative to the [100] axis of the principal axis of the anisotropy responsible for the VBM.

Recombination of one of the electrons with the hole leaves a final state with an electron coupled to the Mn by ${\cal H}_{e-Mn}=I_{e_1Mn}\vec{S}\cdot\vec{\sigma}$. $I_{e_1Mn}$ is the exchange integral of the electron in the ground state with the Mn. The resulting twelve eigenstates of the electron-Mn complex are split into a ground state septuplet (total spin $M=3$) and a higher energy level with $M=2$ (degeneracy of 5). We label them all as $|M,M_z\rangle$. 

For each of the 6 levels of $X^-$-Mn there are 2 possible final states with either $M=2$ or $M=3$ after annihilation of an e-h pair. The Mn spin is not affected by the optical transition and their weight is given by both optical and spin conservation rules. We consider, for instance, a $\sigma+$ transitions where the $|\downarrow_1,\Uparrow_1\rangle$ e-h pair is annihilated. After the recombination, the resulting state is $|S_z,\uparrow_1\rangle$. The intensity of the optical transition from an initial state $|\uparrow_1, \downarrow_1\rangle\times|S_z,\Uparrow_1\rangle$ is proportional to the overlap $\langle M,M_z|S_z,\uparrow_1\rangle$ (Clebsh-Gordan coefficient of the composition of a spin 1/2 with a spin 5/2). 

The highest energy transition corresponds to the high energy initial state $|\uparrow_1, \downarrow_1\rangle\times|+5/2,\Uparrow_1\rangle$ and to the final state $|+5/2,\uparrow_1\rangle$ which is identical to $|3,+3\rangle$ and thereby gives the highest optical weight. In contrast, transition to the final state $|2,M_z\rangle$ which does not contain any $|+5/2,\uparrow_1\rangle$ component is forbidden. The other five doublets have optical weights lying between 1/6 and 5/6 with both $|2,M_z\rangle$ and $|3,M_z\rangle$ final states and the number of resolved lines is 11.

A linear polarization map of the PL of $X^-$-Mn is presented for QD1 in Fig.~\ref{Fig1}(a). Linearly polarized lines are observed on the low energy side of the center of the structure. They arise from spin-flip interaction between the Mn and the hole induced by the presence of VBM \cite{Leger2006}. Provided that $\rho_s/\Delta_{lh}<<1$, the effect of this interaction is small on all the h-Mn doublets except for the states $|-1/2,\Uparrow_1\rangle$ and $|+1/2,\Downarrow_1\rangle$ which are initially degenerated. The bonding and antibonding combinations of these states are coupled, via linearly polarized photons, to the $|2,0\rangle$ and $|3,0\rangle$ electron-Mn states and linearly polarized lines are observed on the PL spectra. Polarization directions are controlled by the strain in the QD plane through the Bir-Pikus Hamiltonian which describes the valence band structure \cite{kp}.

We can obtain numerical values of $I_{h_1Mn}$, $I_{e_1Mn}$ and $\rho_s/\Delta_{lh}$ comparing the transition probabilities calculated from a diagonalization of ${\cal H}_{h-Mn}$ and ${\cal H}_{e-Mn}$ with the experimental data (inset of Fig.\ref{Fig1}(a)). A VBM is observed in most of the QDs but the value of the coefficient $\rho_s/\Delta_{lh}$ shows that the hole-Mn exchange interaction remains highly anisotropic (dominant heavy-hole character). These parameters can be confirmed by an analysis of the magnetic field dependence of the emission. A longitudinal magnetic field dependence is presented in Fig.~\ref{Fig1}(a) and compared with the result of the modeling with the same parameters as the one used to calculate the linear polarization map at zero field.

Under high excitation power, additional emission multiplets usually appear on the low energy side of X$^-$-Mn (see QD2, QD3, QD4 in Fig.~\ref{Fig1}). As we will discuss in section V, they correspond to neutral and charged biexciton interacting with the Mn spin.

\section{Excited states in negatively charged Mn-doped QDs.}

\begin{figure}[hbt]
\centering
\includegraphics[width=1.0\linewidth]{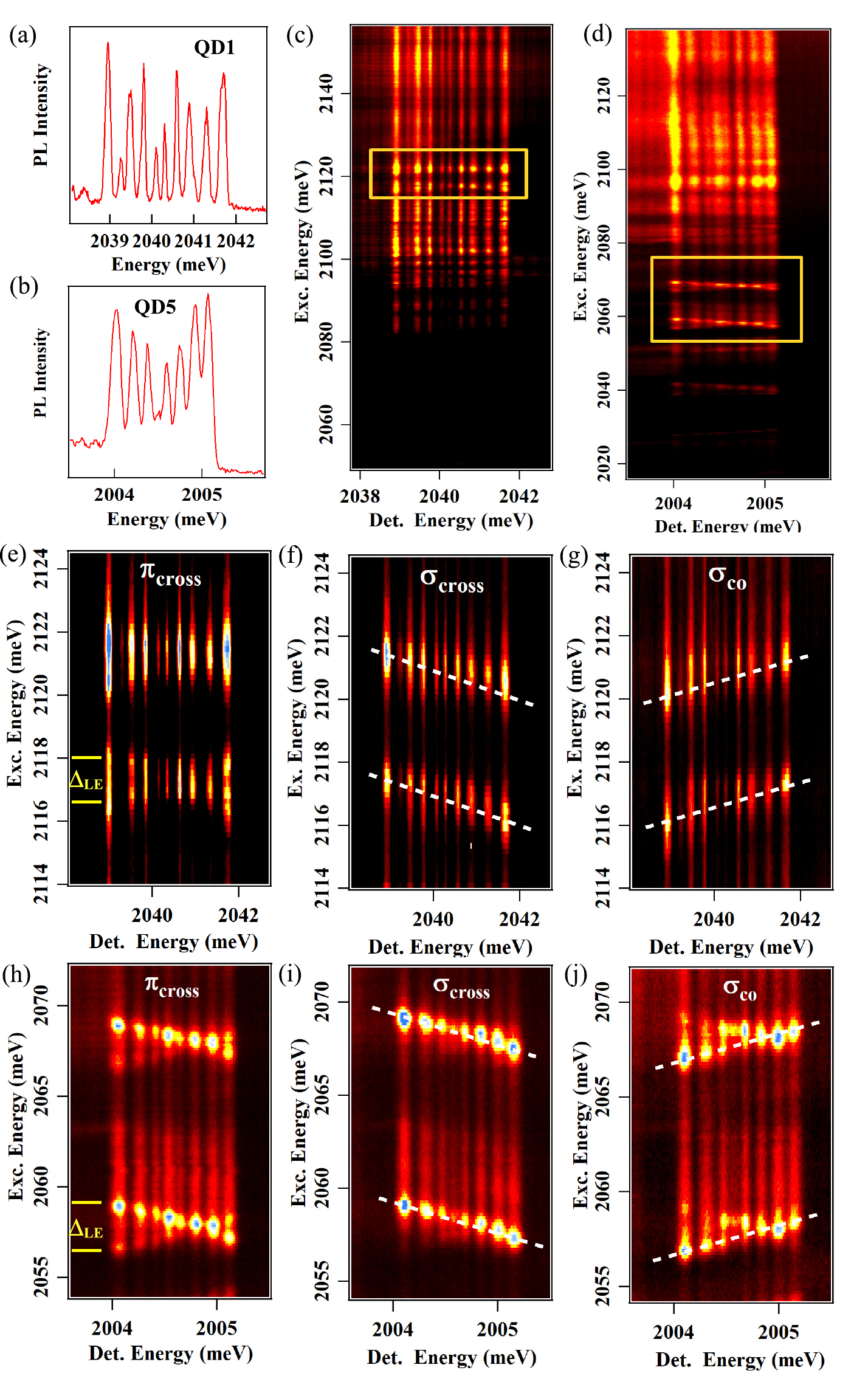}
\caption{Polarized fine structure of the excited states in two negatively charged Mn-doped QDs, QD1 with a large ground state splitting and QD5 with a smaller ground state splitting. (a) PL of X$^-$-Mn in QD1. (b) PL of X$^-$-Mn in QD5. (c)  PLE intensity map in $\pi_{cross}$ excitation/detection configuration in QD1. (d) PLE intensity map in $\pi_{cross}$ excitation/detection configuration in QD5. (e), (f) and (g) detail of PLE of QD1 for the excited states highlighted in (c) in $\pi_{cross}$,  $\sigma_{cross}$ and $\sigma_{co}$ configuration respectively. (h), (i) and (j) detail of PLE of QD5 for the excited states highlighted in (d) in $\pi_{cross}$,  $\sigma_{cross}$ and $\sigma_{co}$ configuration respectively.}
\label{FigPLE}
\end{figure}

Let us first focus on the optical properties of the singly charged exciton. PLE spectra of X$^-$-Mn are presented in Fig.~\ref{FigPLE} for two QDs, QD1 which presents a large ground state exchange splitting and QD5 with a weaker splitting. For both QDs, absorption resonances in the PLE spectra are observed on top of an absorption background. For a cross-linear excitation/detection configuration ($\pi_{cross}$), absorption structure with an asymmetric cross-like shape are observed in the PLE intensity maps (see details in Fig.~\ref{FigPLE}(e) for QD1 and Fig.~\ref{FigPLE}(h) for QD5). Such structures are usually observed for different excited states of the charged magnetic dots. 

When spin-polarized e-h pairs are injected through circularly polarized photo-excitation on excited states, PLE spectra reveals resonances with a strong circular polarization dependence. This is illustrated in Fig.~\ref{FigPLE}(f) and (g) for QD1 and Fig.~\ref{FigPLE}(i) and (j) for QD5. The co-circularly polarized ($\sigma_{co}$) and the cross-circularly polarized ($\sigma_{cross}$) PLE intensity maps clearly show a different excitation energy dependence with a positive slope in $\sigma_{co}$ ({\it i.e.} low energy excitation gives a PL on the low energy (LE) line and high energy excitation gives a PL on the high energy (HE) line) and a negative slope in $\sigma_{cross}$ ({\it i.e.} low energy excitation gives a PL on the HE line and high energy excitation gives a PL on the LE line). The observation of strongly circularly polarized emission (either $\sigma_{co}$ or $\sigma_{cross}$) shows first that the spin of the hole-Mn complex is well conserved during the lifetime of X$^-$. 

\begin{figure}[hbt]
\centering
\includegraphics[width=1.0\linewidth]{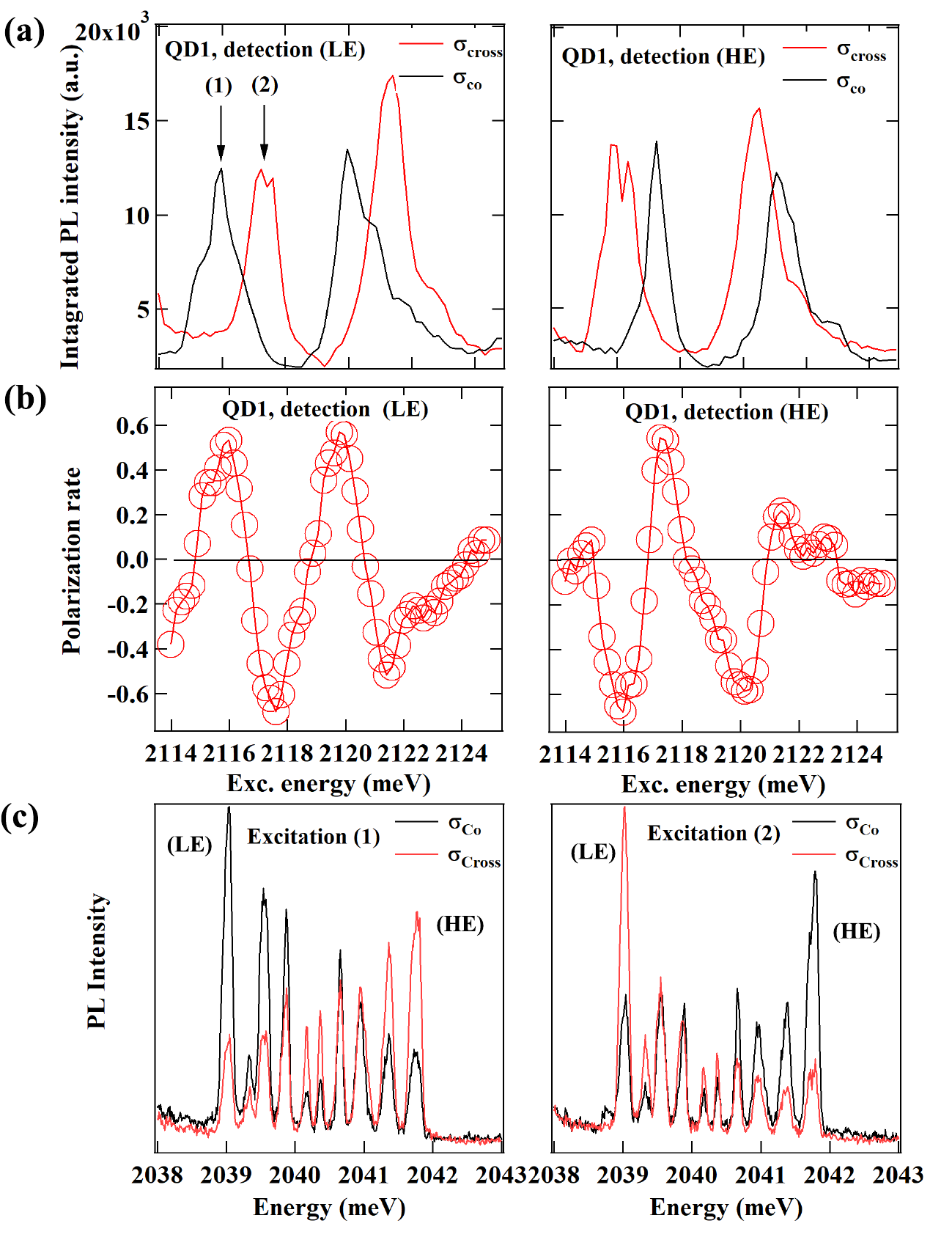}
\caption{(a) Circularly polarized PLE spectra of X$^-$-Mn in 
QD1 for a detection on the low energy line (left) and on the high energy line (right). (b) Corresponding circular polarization rate. (c) Circularly polarized PL spectra obtained for a resonant excitation on (1) (left panel) and on (2) (right panel).}
\label{Fig3}
\end{figure}

For each line of X$^-$-Mn, PLE resonances displaying a fine structure doublet are resolved in the circularly polarized PLE maps. This is particularly clear on the HE and LE lines. For instance, as detailed for QD1 in Fig.~\ref{Fig3}, a switching between positive and negative circular polarization rate is observed as the laser excitation energy is increased around the excited states at 2116 meV and 2121 meV. 

The splitting of the observed absorption doublet depends on the X$^-$-Mn line. In particular, the order of the circularly polarized absorption lines is reversed for a detection on HE and LE lines (Fig.~\ref{Fig3}). Whereas for a detection on LE, a $\sigma_{co}$ emission is first observed when the excitation energy is increased, the order is reversed for a detection on HE where the $\sigma_{cross}$ PL appears first. For a detection in the center of the PL spectra, the two absorption resonances are usually not resolved in the PLE spectra.

Such doublet structure with a circular polarization reversal can usually be observed on different excited states of X$^-$-Mn. The measured splitting is changing from an excited state to another and from dot to dot. This is illustrated for instance by a comparison of exited states in QD1 and QD5. Whereas the splitting between the $\sigma_{co}$ and the $\sigma_{cross}$ PLE resonances measured on the low energy line $\Delta_{LE}\approx 1.4 meV$ on QD1 (Fig.~\ref{FigPLE}(e)), a much larger value is observed on QD5 with $\Delta_{LE}\approx 2.2 meV$ (Fig.~\ref{FigPLE}(h)). QD5 with the smallest carriers-Mn exchange interaction in the ground state of X$^-$-Mn presents the largest splitting in the excited states.

\subsection{The triplet states of the charged exciton coupled with a Mn spin.}

To understand this circularly polarized structure in the PLE, one has to consider that the optical injection of an e-h pair on an excited state of a negatively charged QD creates an excited negatively charged exciton, X$^{-*}$-Mn. It consists in a resident electron in the ground state of the dot and an e-h pair in the resonantly addressed excited state. 

\begin{figure}[!hbt]
\begin{center}
\includegraphics[width=1.0\linewidth]{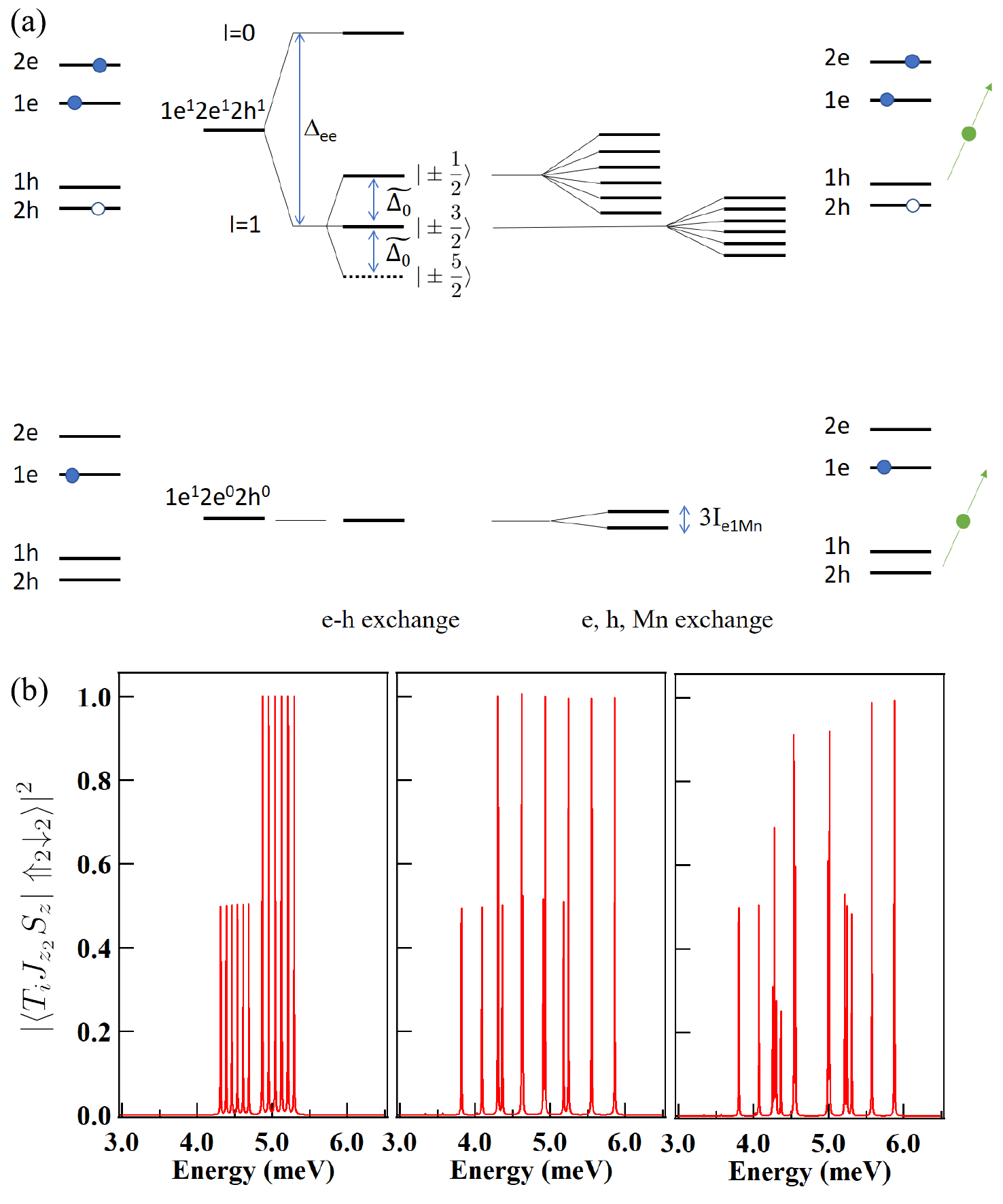}
\caption{(a) Scheme of the energy levels of a resonantly created X$^{-*}$ without (left) and with (right) exchange coupling with a Mn spin. (b) Calculated energy levels of the X$^{-*}$ triplet states interacting with a Mn spin. The intensity gives the bright exciton ($\sigma+$) component and only the bright triplets appear. The parameters used in the calculation are: $I_{e_2Mn}$=0 $\mu eV$, $\tilde{\Delta_{ee}}$=-5000 $\mu eV$, $\tilde{\Delta_{0}}$=-580 $\mu eV$, $\tilde{\Delta_{1}}$=0 $\mu eV$, $\rho_s/\Delta_{lh}$= 0, $\theta_s$= 0 and (left) $I_{e_1Mn}$=-30$\mu eV$, $I_{h_2Mn}$=50 $\mu eV$, (center) $I_{e_1Mn}$=-80$\mu eV$, $I_{h_2Mn}$=180 $\mu eV$ and (right) $I_{e_1Mn}$=-80$\mu eV$, $I_{h_2Mn}$=180 $\mu eV$ and $\tilde{\Delta_{1}}$=250 $\mu eV$.}
\label{schemeX-}
\end{center}
\end{figure}

For a sufficiently large confinement of the carriers on the excited state, the e-e and the e-h exchange interaction induce a fine-structure of X$^{-*}$-Mn (Fig.~\ref{schemeX-}). In a non-magnetic dot, the strongest term is the e-e exchange interaction which reduces in a spherical approximation to a Heisenberg Hamiltonian \cite{Urbaszek2003}:

\begin{eqnarray}
H_{e_1,e_2}=\Delta_{ee} \vec{\sigma}_1.\vec{\sigma}_{2}
\end{eqnarray}

\noindent where $\vec{\sigma}_1$ ($\vec{\sigma}_{2}$) is the spin operator of the electron on the ground (excited) state and $\Delta_{ee}<0$ the exchange integral. This Hamiltonian splits the excited singlet of the two electrons $S^*_0$ with total angular momentum $I=0$ (where $I=\sigma_{1}+\sigma_{2}$) from the lower energy triplet states $T_{0}$, $T_{\pm 1}$ with $I=1$ and I$_z$=0,$\pm1$. This splitting is typically equal to a few $meV$ for an electron on an excited state of InAs/GaAs self-assembled QDs \cite{Cortez2002, Ware2005} and could be larger in II-VI compounds presenting a larger Coulomb interaction.

Within X$^{-*}$, the e-h exchange interaction $\Delta_0$ is a smaller correction (typically a few $100\mu eV$ \cite{Legall2012}) which splits the triplet states (see Fig.~\ref{schemeX-}(a)). The two electrons are coupled to each other much strongly than each of them with the hole. One can then consider the exchange interaction of the three particles as an interaction of the hole spin with the total spin of the two electrons $I=\sigma_{1}+\sigma_{2}$. This e-h exchange interaction can be written in a compact form \cite{Kavokin2003}:

\begin{eqnarray}
H_{e_{1}e_{2}h_{2}}=2\tilde{\Delta}_0I_z\sigma^h_z+\tilde{\Delta}_1(I_x\sigma^h_x+I_y\sigma^h_y)
\label{exceff}
\end{eqnarray}

\noindent with $\sigma^h_i=1/2\tau_i$ (with $\tau_i$ the Pauli matrices) acting on the heavy-hole subspace. $\tilde{\Delta_{0,1}}=1/2(\Delta_{0,1}^{1}+\Delta_{0,1}^{2})$ is an average of the hole exchange interaction with the electron on the ground ($\Delta_{0,1}^{1}$) and excited ($\Delta_{0,1}^{2}$) state. The second term in (\ref{exceff}) is an effective description of the long range part of the e-h exchange interaction which becomes non-zero in an anisotropic confinement potential \cite{Bayer2002,Leger2007}. 

In the case of a simple e-h pair in the QD ground state, this effective spin Hamiltonian with $\tilde{\Delta_0}<$0 stabilizes the states with parallel electron and hole spins (split the high energy radiative and the low energy non-radiative excitons) and split the high energy radiative exciton with the energy $\tilde{\Delta_1}$ (usual fine structure splitting of the exciton).

In the case of X$^{-*}$, this e-h exchange interaction splits the electron triplet states and the charged exciton states read \cite{Akimov2005}:

\begin{eqnarray}
\vert\pm\frac{1}{2}\rangle=&T_{\mp1}\vert\pm\frac{3}{2}\rangle_{2}=\left(\vert\mp\frac{1}{2}\rangle_{1}\vert\mp\frac{1}{2}\rangle_{2} \right) \vert\pm\frac{3}{2}\rangle_{2} \nonumber \\
\vert\pm\frac{3}{2}\rangle=& T_0\vert\pm\frac{3}{2}\rangle_{2}\nonumber\\
=&\frac{1}{\sqrt{2}}\left(\vert+\frac{1}{2}\rangle_{1}\vert-\frac{1}{2}\rangle_{2}+\vert-\frac{1}{2}\rangle_{1}\vert+\frac{1}{2}\rangle_{2}\right)\vert\pm\frac{3}{2}\rangle_{2} \nonumber \\
\vert\pm\frac{5}{2}\rangle=&T_{\pm1}\vert\pm\frac{3}{2}\rangle_{2}=\left(\vert\pm\frac{1}{2}\rangle_{1}\vert\pm\frac{1}{2}\rangle_{2} \right) \vert\pm\frac{3}{2}\rangle_{2} 
\end{eqnarray}

\noindent where the X$^{-*}$ states are labelled with their total angular momentum projections along the QD growth axis z. The non-diagonal exchange terms $\tilde{\Delta_1}$ couples the triplet states $T_{-1}\vert+\frac{3}{2}\rangle_{2}$ and the states $T_{0}\vert-\frac{3}{2}\rangle_{2}$ on one side and  $T_{+1}\vert-\frac{3}{2}\rangle_{2}$ and   $T_{0}\vert+\frac{3}{2}\rangle_{2}$  on the other side. In the absence of VBM and in the absence of interaction with a magnetic atom, the $\vert\pm\frac{5}{2}\rangle$ triplet states have parallel electron and hole spins in the excited state and cannot be created by a resonant optical excitation.  

In a Mn-doped QD, these split X$^{-*}$ states interact with the spin of the magnetic atom. This can be described by the carriers-Mn spin effective Hamiltonian:

\begin{eqnarray}
H_{X^{-*}-Mn}=H_{e_{1}e_{2}h_{2}}+\nonumber\\
I_{e_1Mn}\vec{\sigma_1}\vec{S}+I_{e_{2}Mn}\vec{\sigma_{2}}\vec{S}+I_{h_{2}Mn}\vec{J_{2}}\vec{S}
\end{eqnarray}

\noindent where $I_{e_iMn}$ ($I_{h_iMn}$) are the exchange integrals between the electrons (hole) and the Mn spins. The exchange integrals for carriers on excited states depend on the overlap with the magnetic atom and can significantly change from an excited state to another. As for carriers confined in the ground state, the exchange interaction is expected to be dominated by the anti-ferromagnetic coupling with the hole spin. It can also be affected by a possible VBM. 

Calculated energy levels of the triplets states of X$^{-*}$ interacting with a Mn spin are presented in Fig.~\ref{schemeX-}(b). If the exchange interaction with the Mn spin is weaker than the e-h exchange interaction $\tilde{\Delta_0}$, it further splits into six levels the optically active states of X$^{-*}$ (Fig.~\ref{schemeX-}(b), left panel). The largest splitting is obtained for the highest energy triplet states $\vert\pm1/2\rangle$ where the anti-ferromagnetic interaction with the hole and ferromagnetic interaction with both electrons add. The splitting of the lower energy states $\vert\pm3/2\rangle$ is only controlled by the exchange interaction with the hole spin on the excited state. The lowest energy triplet is dark and do not appear in Fig.~\ref{schemeX-}(b) which only displays a bright exciton component $\vert\langle T_i J_{z_2} S_z\vert \Uparrow_2 \downarrow_2\rangle\vert^2$. 

The two bright triplets can overlap if the energy shift induced by the exchange interaction with the Mn is larger than the e-h exchange splitting $\tilde{\Delta_0}$ (Fig.~\ref{schemeX-}(b), center panel). In the presence of an anisotropic e-h exchange interaction a gap appears in the center of the structure (Fig.~\ref{schemeX-}(b), right panel). The width of the gap is controlled by $\tilde{\Delta_1}$. This mixing can induce linear polarization in the absorption of the triplets with polarization directions controlled by the shape anisotropy of the wave function in the excited state. Neglecting this possible mixing, each level of the split triplet corresponds to a given spin state S$_z$ of the Mn.

\subsection{Circularly polarized fine structure of the excited states in a charged Mn-doped QD.}

The optical excitation of an excited state of a charged dot creates a triplet state of X$^{-*}$-Mn. Independently of its spin, the optically created hole can relax towards the ground state of the dot. For the electron, the relaxation channels are not straightforward and depends on the relative orientation of the resident and injected electrons' spin. For a given Mn spin state S$_z$, the relaxation of X$^{-*}$-Mn towards X$^-$-Mn will then depend on the triplet state which is created. 
 
\begin{figure}[hbt]
\centering
\includegraphics[width=1.0\linewidth]{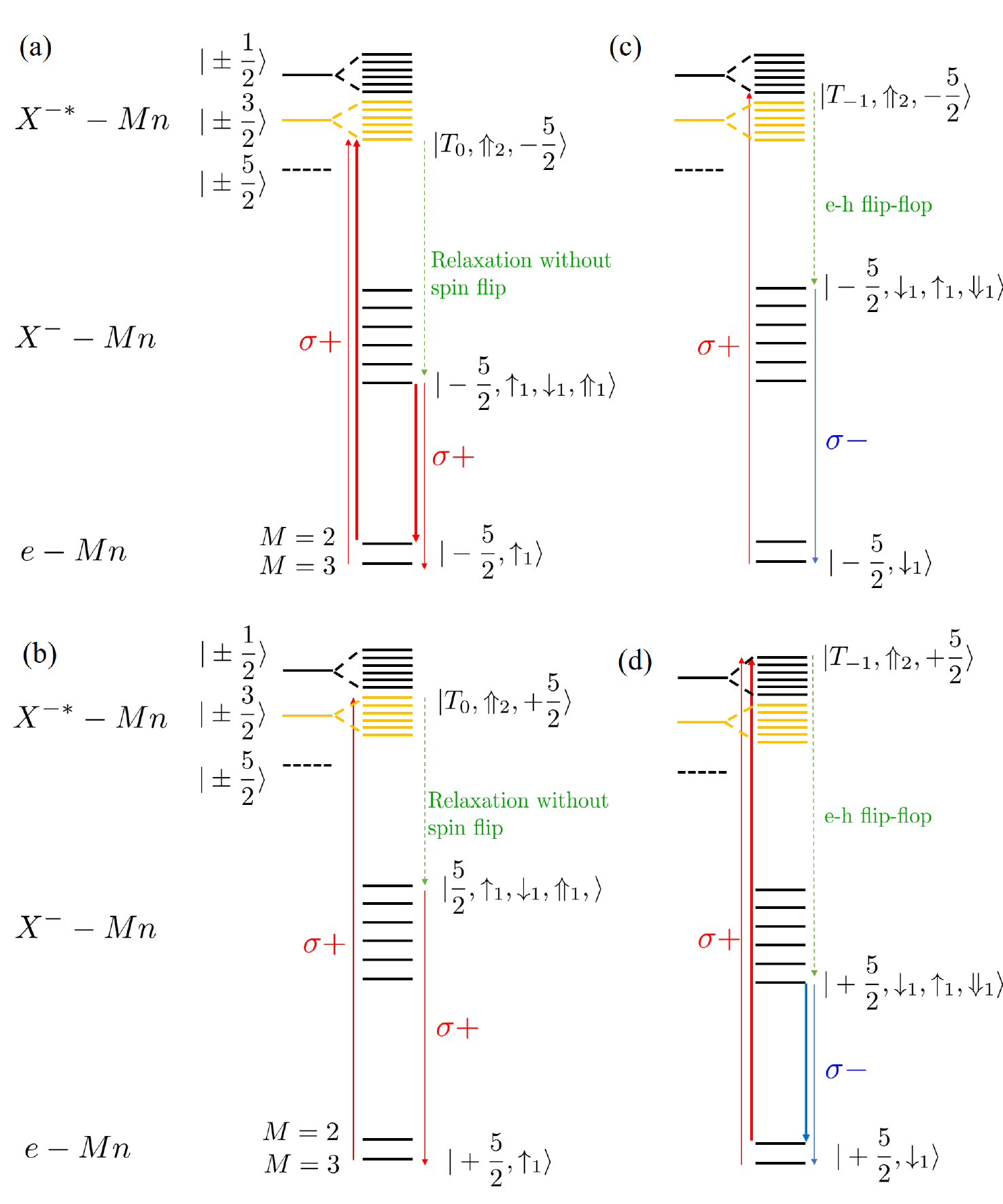}
\caption{Path of resonant optical excitation and spin relaxation in a negatively charged Mn-doped QD which contains a spin up electron ((a) and (b)) or a spin down electron ((c) and (d)). Only the case of resonant excitation on the low and the high energy levels of the split triplet states are presented.}
\label{Figpath}
\end{figure}

Let's consider for instance a $\sigma+$ excitation which creates an e-h pair $\vert\downarrow_2,\Uparrow_2\rangle$ on an excited state of a charged magnetic dot. This e-h pair is exchanged coupled with the electron resident on the ground state and the embedded Mn atom. The exchange interaction with the Mn spin is dominated by the anti-ferromagnetic coupling with the spin up hole $\vert\Uparrow_2\rangle$. In a first approximation, the lowest energy triplet states $\vert\pm\frac{5}{2}\rangle$ are dark and cannot produce any significant resonance in the PLE spectra.

When a spin up electron $\vert\uparrow_1\rangle$ is present in the dot, an absorption can occur on the triplet state $\vert+\frac{3}{2}\rangle$. The lowest energy triplet state $\vert T_0,\Uparrow_2,-\frac{5}{2}\rangle$ is associated with S$_z$=-5/2 (Fig.~\ref{Figpath}(a)). This state overlaps with the two electron-Mn initial states with M=2 or M=3 and two optical transitions are possible. The intensity of the transitions are respectively proportional to $\vert\langle2,-2\vert-\frac{5}{2},\uparrow_1\rangle\vert^2$=5/6 and $\vert\langle3,-2\vert-\frac{5}{2},\uparrow_1\rangle\vert^2$=1/6. This gives rise to a large intensity transition from M=2 and a higher energy less intense transition from M=3. As the resident and created electrons have anti-parallel spins, the injected exciton can relax to the ground state with a conservation of the spin of both the electron and the hole. This relaxation does not involve an interaction with the Mn spin which is expected to be conserved. This triplet state relaxes to the low energy X$^-$-Mn level $\vert-\frac{5}{2}\uparrow_1,\downarrow_1,\Uparrow_1\rangle$ without any spin-flip and recombines emitting a $\sigma+$ photon ({\it i.e.} co-polarized with the excitation) on the low energy line of X$^-$-Mn (final state M=2) or at slightly higher energy on a less intense transition towards the level M=3.

For a spin state of the Mn S$_z$=+5/2 and a resident spin up electron $\vert\uparrow_1\rangle$, the absorption takes place on the high energy level of the $\vert+3/2\rangle$ bright triplet $\vert T_0,\Uparrow_2,+\frac{5}{2}\rangle$ (Fig.~\ref{Figpath}(b)). This state can also relax to X$^-$-Mn without any spin-flip. It ends up on the states $\vert+\frac{5}{2},\uparrow_1,\downarrow_1,\Uparrow_1\rangle$ and produces a co-polarized ($\sigma+$) emission on the high energy line of X$^-$-Mn. 

With an increase of the excitation energy, one successively excites the different spin states of the Mn giving rise to successive co-polarized emission on the different X$^-$-Mn lines starting from the low energy side to the high energy side of the X$^-$-Mn spectra.

Lets now consider that the resident electron is spin down, $\vert\downarrow_1\rangle$. For a $\sigma+$ excitation which creates an e-h pair $\vert\downarrow_2,\Uparrow_2\rangle$, an absorption cannot occur on the triplet states $\vert\pm3/2\rangle$ as they do not contain components with parallel electron spins. An absorption can take place at higher energy on the charged exciton states $\vert+\frac{1}{2}\rangle$ associated with the electron triplet T$_{-1}$.

The exchange interaction with the magnetic atom is still dominated by the anti-ferromagnetic coupling with the spin up hole $\vert\Uparrow_2\rangle$ and the lowest energy state that can be excited corresponds to S$_z$=-5/2 (state $\vert T_{-1},\Uparrow_2,-\frac{5}{2}\rangle$ in Fig.~\ref{Figpath}(c)). Now, the resident and created electrons have parallel spins (both spin down in this case) and the relaxation to X$^-$ from T$_{-1}$ $\vert\Uparrow_2\rangle$ cannot occur directly. The  relaxation requires an e-h flip-flop to the state T$_{0}$ $\vert\Downarrow_2\rangle$ before relaxing to the X$^-$ singlet state S$_0$ $\vert\Downarrow_1\rangle$ \cite{Cortez2002,Legall2012}. The Mn spin is not involved in this relaxation process. It is conserved and the final state after relaxation, $\vert-\frac{5}{2},\downarrow_1,\uparrow_1,\Downarrow_1\rangle$, has parallel hole and Mn spins. This corresponds to the highest energy X$^-$-Mn state which recombines emitting a $\sigma-$ photon ({\it i.e} cross-polarized with the resonant excitation) on the high energy line (Fig.~\ref{Figpath}(c)). 
 
For a spin state of the Mn S$_z$=+5/2 and a spin down resident electron $\vert\downarrow_1\rangle$, a $\sigma+$ excitation creates the state $\vert T_{-1},\Uparrow_2,+\frac{5}{2}\rangle$, the high energy level associated with the triplet $\vert+\frac{1}{2}\rangle$. Two optical transitions are possible, from M=2 and M=3 (Fig.~\ref{Figpath}(d)). An e-h flip-flop is also required for the relaxation toward the low energy X$^-$-Mn state $\vert+\frac{5}{2},\downarrow_1,\uparrow_1,\Downarrow_1\rangle$. It recombines on the two low energy lines of X$^-$-Mn emitting a $\sigma-$ photon.

In this case, an increase of the excitation energy gives rise to successive cross-circularly polarized emission starting from the high energy to the low energy line of X$^-$-Mn PL spectra. This series of absorption is shifted towards high energy compared to the $\sigma_{co}$ structure by the isotropic part of the e-h exchange interaction $\tilde{\Delta_0}$.

These two series of absorption are at the origin of the circularly polarized cross-like structure observed in the PLE spectra of X$^-$-Mn (Fig.~\ref{FigPLE}). For a given emission line of X$^-$-Mn, we can assign the observed doublets in the PLE spectra to the excitation of the two bright triplet states of X$^{-*}$. The relative energy positions of the absorption lines giving rise to the $\sigma_{co}$ and $\sigma_{cross}$ PL are controlled both by the e-h exchange interaction $\tilde{\Delta_0}$ and the exchange coupling with the spin of the magnetic atom.

\begin{figure}[hbt]
\centering
\includegraphics[width=1.0\linewidth]{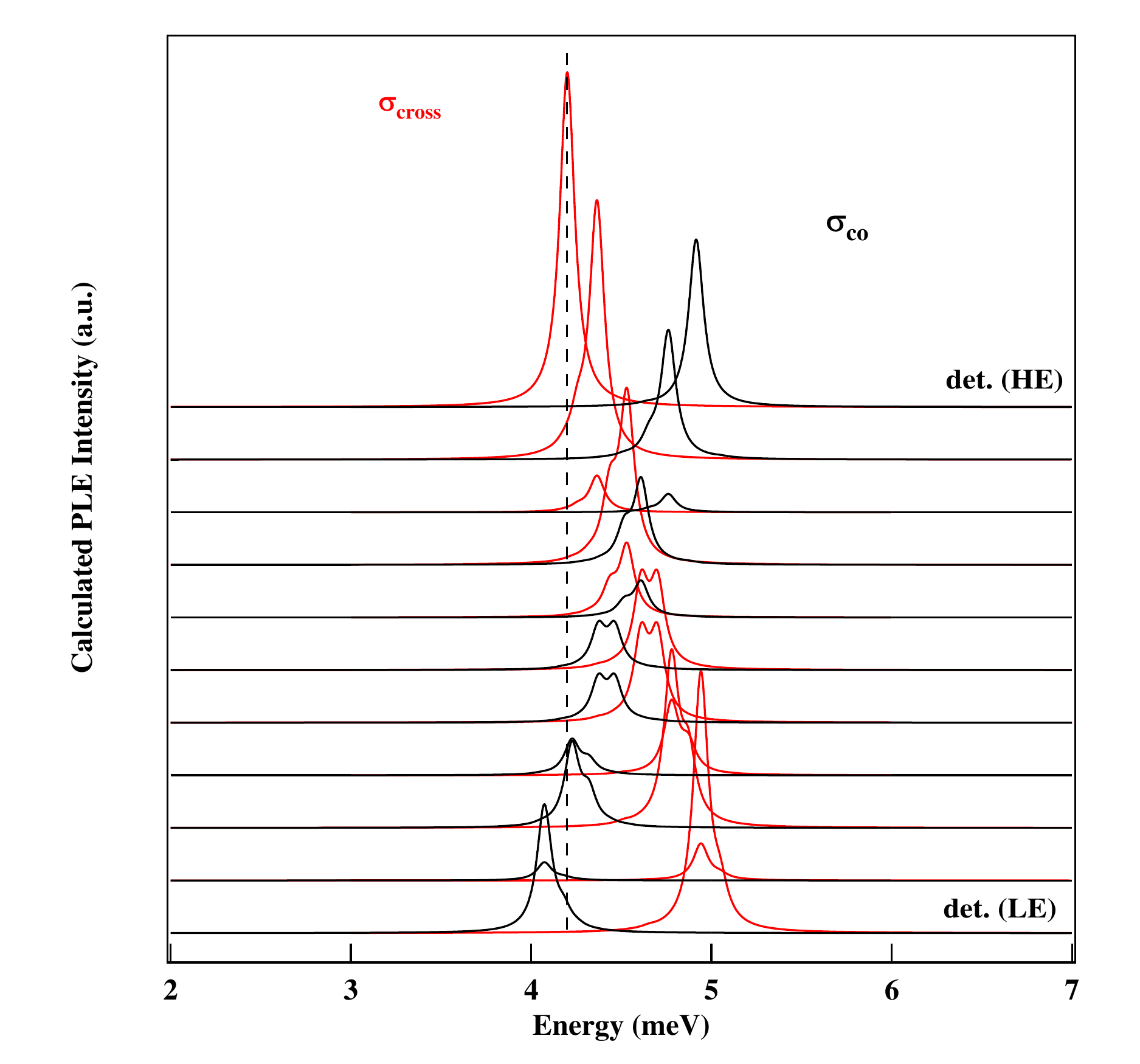}
\caption{Calculated PLE spectra of X$^-$-Mn for a QD containing a spin up electron $\vert\uparrow_1\rangle$. PLE of the 11 emission lines of X$^-$-Mn are displayed for a $\sigma+$ detection and a $\sigma+$ excitation (red) or $\sigma-$ excitation (black). Spectra are shifted for clarity and a line broadening with a Laurentzian of half width at half maximum of 50$\mu eV$ is included. The bottom trace corresponds to a detection on the low energy line (LE) and the top trace to a detection on the high energy line (HE). The parameters used in the calculations are $I_{e_1Mn}$= -30$\mu eV$, $I_{e_2Mn}$= 0$\mu eV$, $I_{h_2Mn}$= 100$\mu eV$, $\tilde{\Delta_{0}}$= -75$\mu eV$, $\tilde{\Delta_{1}}$= 0$\mu eV$, $\Delta_{ee}=-5000 \mu eV$, $\rho_s/\Delta_{lh}$= 0, $\theta_s$= 0. These values reproduce the order of magnitude of the splitting and shift observed on the triplet states in QD1.}
\label{FigMod}
\end{figure}

Calculated PLE spectra of X$^-$-Mn are presented in Fig.~\ref{FigMod}. In this model we consider that the QD contains a spin up electron $\vert\uparrow_{1}\rangle$ (identical results are obtained with a spin down electron by reversing all the circular polarizations). For an estimation of the PLE spectra we also consider that the spin of the Mn and of the resident electron are conserved during the optical transitions and during the relaxation of the optically created e-h pair from an excited state of the dot to the ground state of the charged exciton.

To obtain the PLE signal, we first calculate the absorption amplitude of the transition between the e-Mn levels in the ground state and each triplet state of the charged exciton in the excited state. The amplitude of transition is calculated in each circular polarization. For instance, for a resident electron spin up $\vert\uparrow_1\rangle$, the amplitude of the $\sigma+$ transitions which creates an e-h pair $\vert\downarrow_2,\Uparrow_2\rangle$ is given by the overlap $\langle M,M_z\vert S_z,\uparrow_1,\downarrow_2,\Uparrow_2\rangle$. This corresponds to an excitation of the triplet state T$_0\vert\Uparrow_2\rangle$.

For a given line of X$^-$-Mn, the PL is then given by the product of the intensity of the considered transition $\vert\langle M,M_z\vert S_z,\uparrow_1\rangle\vert^2$ by the probability of the absorption which creates an excited charged exciton state $\vert S_z,\uparrow_1, \downarrow_2, \Uparrow_2\rangle$ for a $\sigma+$ excitation or $\vert S_z,\uparrow_1, \uparrow_2, \Downarrow_2\rangle$ for a $\sigma-$ excitation. The first one gives rise to a $\sigma_{co}$ emission whereas the second one corresponds to a $\sigma_{cross}$ PL. 
 
In the results presented in Fig.\ref{FigMod} an absorption series with a positive slope ({\it i.e.} low energy excitation gives a PL on the low energy line and high energy excitation gives a PL on the high energy line) produces a $\sigma_{co}$ emission and a second series of absorption with a negative slope, slightly shifted to higher energy gives rise to $\sigma_{cross}$ PL. This corresponds to the situation observed in the experiments displayed in Fig.~\ref{FigPLE} and Fig.~\ref{Fig3}. In the experiments, the broadening of the line prevents the observation of the detailed structure of the calculated excitation spectra but a good overall agreement with the model is observed (Fig.~\ref{FigMod}). 

The anisotropic part of the e-h exchange interaction $\tilde{\Delta_1}$ can mix bright triplet states associated with the same Mn spin state. This can introduce some deviation from the purely circularly polarized optical selection rules and induce a perturbation or an energy gap in the center of the absorption structure.

\section{Negatively charged biexciton coupled with a Mn spin}

\begin{figure}[hbt]
\centering
\includegraphics[width=1.0\linewidth]{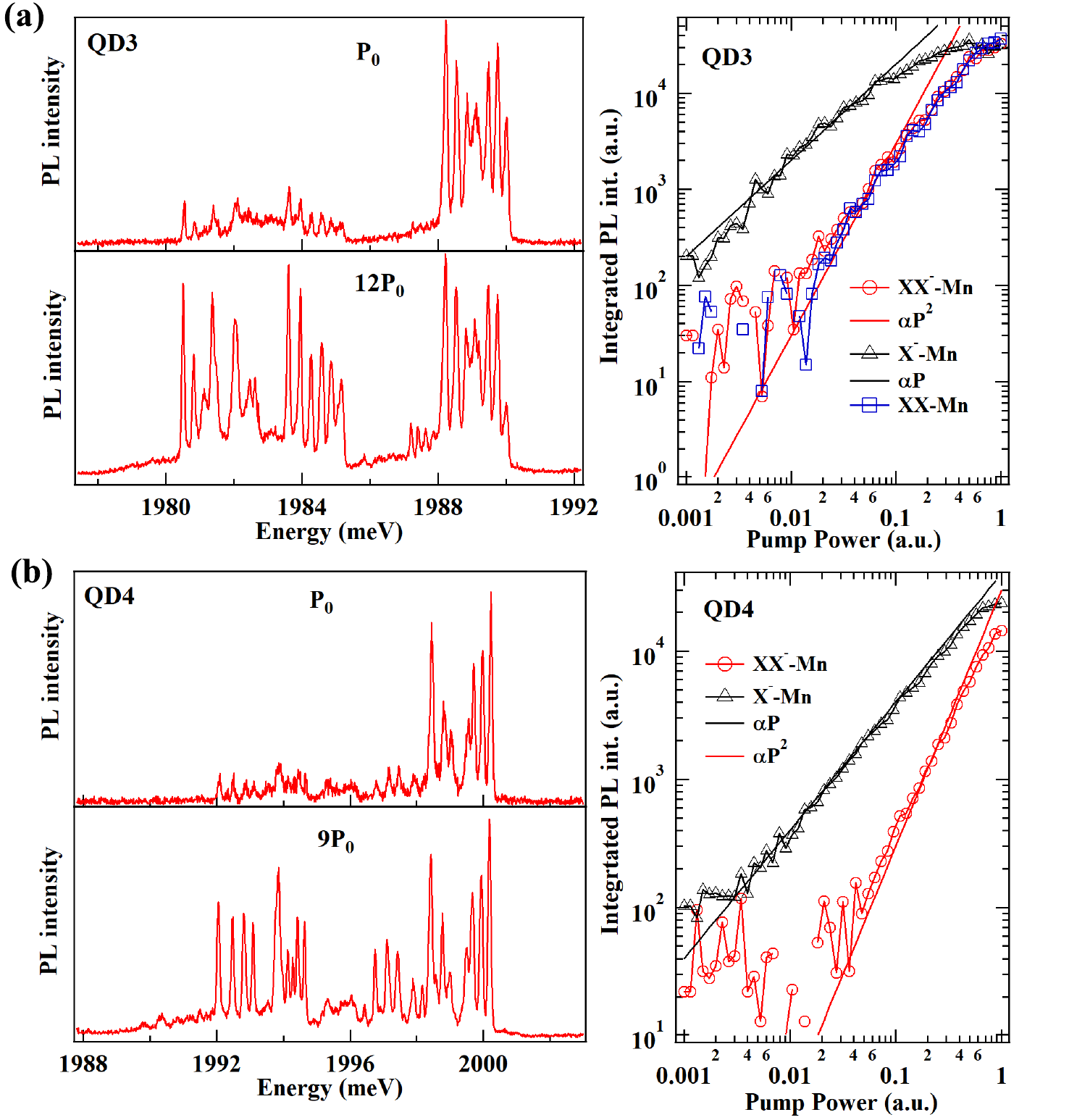}
\caption{(a) Excitation power dependence of the PL of QD3. Left: PL spectra for two excitation powers. Right: power dependence of the integrated PL intensity of X$^-$-Mn, XX-Mn and XX$^-$-Mn. (b) Power dependence of the PL of QD4. Left: PL spectra for two excitation powers. Right: power dependence of the integrated PL intensity of X$^-$-Mn and XX$^-$-Mn. Solid lines correspond to linear (black) and quadratic (red) fits.}
\label{FigXXm}
\end{figure}

\begin{figure}[hbt]
\centering
\includegraphics[width=1.0\linewidth]{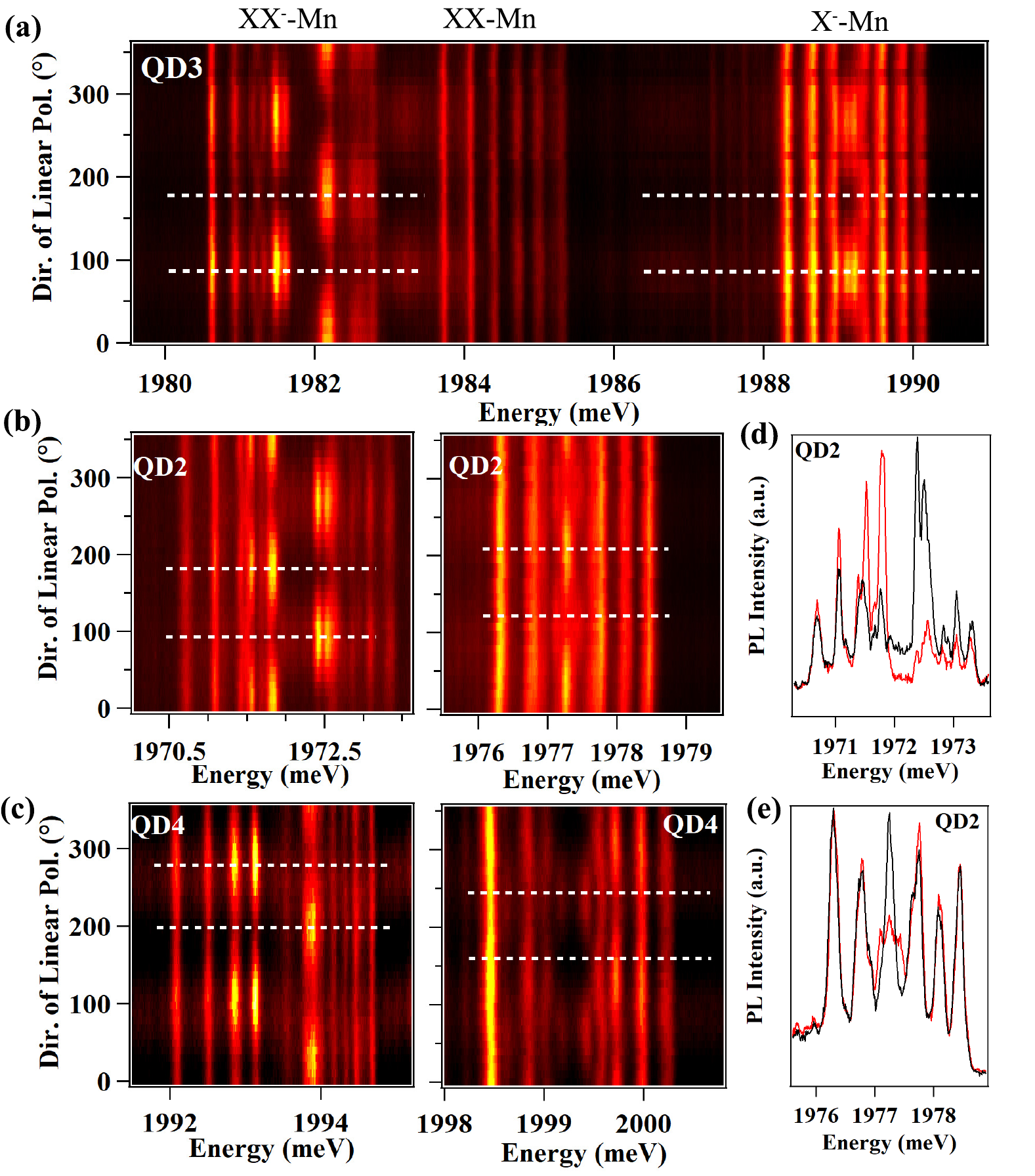}
\caption{(a) Linear polarization properties of X$^-$-Mn, XX-Mn and XX$^-$-Mn in QD3. The direction of polarization are measured with respect to the [100] or [010] axis (45$^{\circ}$ from the cleavage axis of a non-oriented substrate). (b) Linear polarization properties of the PL of XX$^-$-Mn (left) and X$^-$-Mn (right) in QD2. (c) Linear polarization properties of the PL of XX$^-$-Mn (left) and X$^-$-Mn (right) in QD4. (d) and (e), linearly polarized PL spectra of, respectively, XX$^-$-Mn and X$^-$-Mn in QD2. The directions of linear polarization correspond to the doted lines in (b).}
\label{Fig7bis}
\end{figure}

The interaction of the triplet states of an excited charged exciton with the Mn spin can also be observed in the emission of a negatively charged biexciton. In charged magnetic QDs under high excitation power, additional emission lines appear on the low energy side of X$^-$-Mn. The neutral biexciton (XX-Mn) is easily identified with its six lines PL structure and its quadratic excitation power dependence \cite{Besombes2005,Trojnar2011}. However, as presented in Fig.~\ref{FigXXm} for QD3 and QD4, more complex structures consisting in two groups of lines separated by a central gap can also appear. The groups of lines are partially linearly polarized along two orthogonal directions (Fig.~\ref{Fig7bis}). The polarization rate decreases from the center to the outside of the PL structure. The excitation power dependence of their PL intensity is identical to the one observed for the biexciton and is quadratic at low excitation power (Fig.~\ref{FigXXm}). 

This PL structure corresponds to a recombination from the ground state of a negatively charged biexciton interacting with the magnetic atom (XX$^-$-Mn). XX$^-$ consists of two holes and three electrons \cite{Akimov2005}. Within this complex, the angular momenta of the holes as well as the spins of the two electrons which are in the ground state of the dot are compensated. The charged biexciton ground state is then twice degenerate with respect to the spin of the third electron ($\vert\uparrow_1\rangle$ or $\vert\downarrow_1\rangle$) occupying an excited state of the dot.

\begin{figure}[hbt]
\centering
\includegraphics[width=1.0\linewidth]{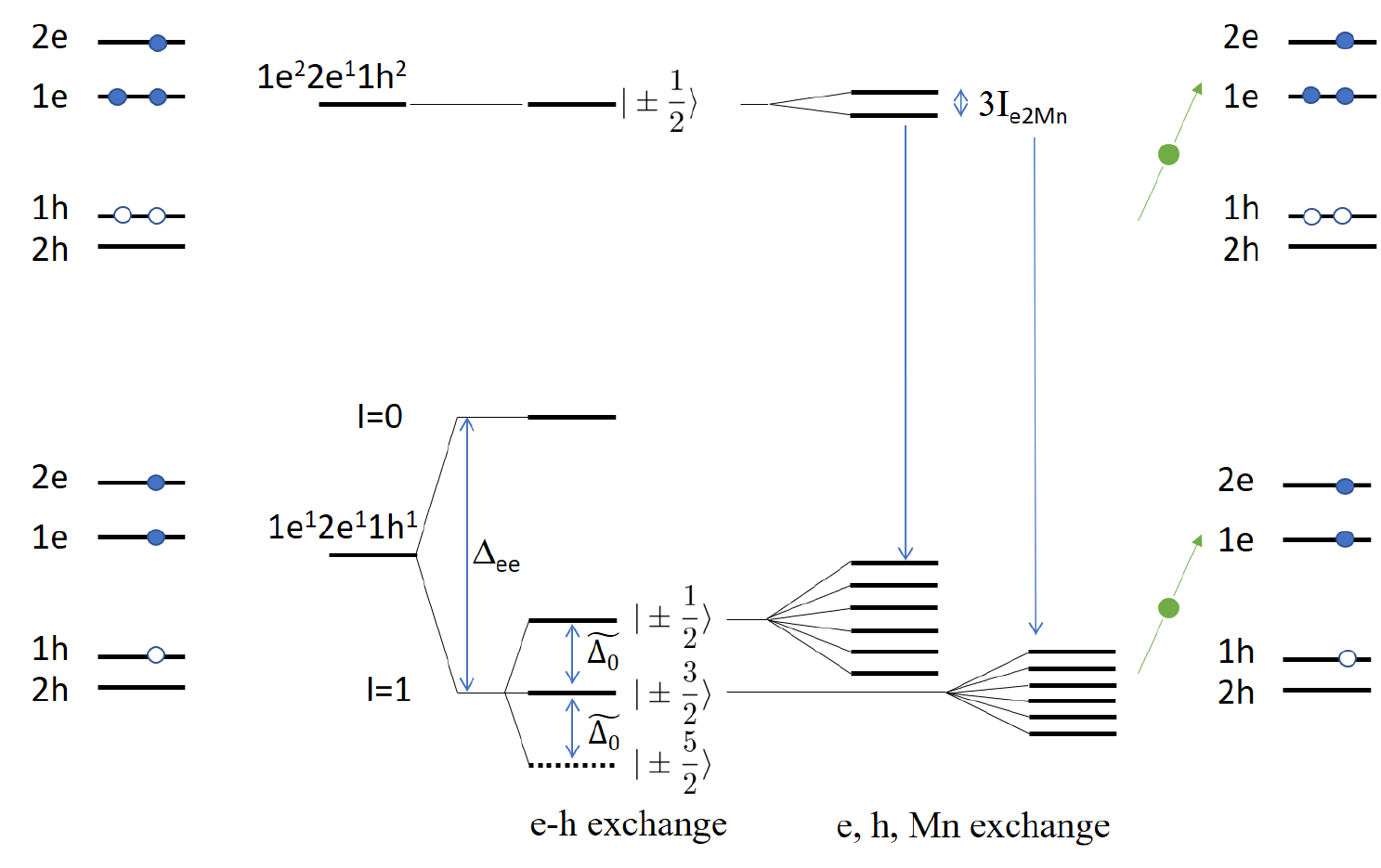}
\caption{Scheme of the energy levels and optical recombination path of XX$^-$ (left) and XX$^-$-Mn (right).}
\label{FigXXmScheme}
\end{figure}

The PL of XX$^-$ arises from the recombination of an e-h pair in the ground state of the dot. An excited charged exciton with an electron in an excited state and an e-h pair in the ground state is left behind as the final state of the radiative recombination (Fig.~\ref{FigXXmScheme}). This excited charged exciton is different from the one created by a resonant excitation on an excited state as now the hole is in the ground state of the QD. It is, however, similarly split into a high energy singlet and a lower energy triplet by the e-e exchange interaction. The triplet states are also split by the e-h exchange interaction. The e-e exchange interaction involves an electron in the ground state and one in an excited state and is identical to the one we discussed for the PLE of X$^-$-Mn. The e-h exchange interaction with the hole which is now more strongly confined in the ground state of the dot is expected to be larger.

In non-magnetic QDs, the recombination of XX$^-$ towards the two bright triplet states of X$^{-*}$ give rise to a PL doublet split by the isotropic part of the e-h exchange interaction $\tilde{\Delta_0}$ \cite{Akimov2005}. In the presence of an anisotropic confinement potential, these optically active triplet states are coupled by the energy $\tilde{\Delta_1}$ and become partially linearly polarized. In the presence of large VBM, the low energy triplet state is not completely dark and transition towards $\vert\pm\frac{5}{2}\rangle$ in the charged biexciton PL could become weakly allowed.

In a magnetic QD, the spin of the electron in the excited state, $\vec{\sigma}_{2}$, is exchange coupled with the spin of the magnetic atom $\vec{S}$ (Fig.~\ref{FigXXmScheme}). As in the case of X$^-$-Mn described in section III, the isotropic coupling between the electron spin 1/2 and the spin 5/2 of the atom should result in two energy levels with total angular momentum M=2 and M=3. The splitting is controlled by the exchange interaction of the electron in the excited state with the Mn spin, $I_{e_2Mn}$. Because of the weaker confinement in the excited state, a value $I_{e_2Mn}<I_{e_1Mn}$ is expected.

\subsection{Polarized fine structure of the negatively charged biexciton}

In the final state of the optical recombination of XX$^-$-Mn, each triplet state of X$^{-*}$-Mn interacts through the exchange interaction with the Mn atom. This interaction is dominated by the anti-ferromagnetic coupling with the heavy-hole in the QD ground state and, for each triplet state, results in a splitting into six energy levels. The energy shift induced by this exchange interaction compete with the anisotropic exchange interaction term $\tilde{\Delta_1}$. $\tilde{\Delta_1}$ mix the two bright triplet states, can induce a gap in the center of the PL structure and a linear polarization rate.

\begin{figure}[hbt]
\centering
\includegraphics[width=1.0\linewidth]{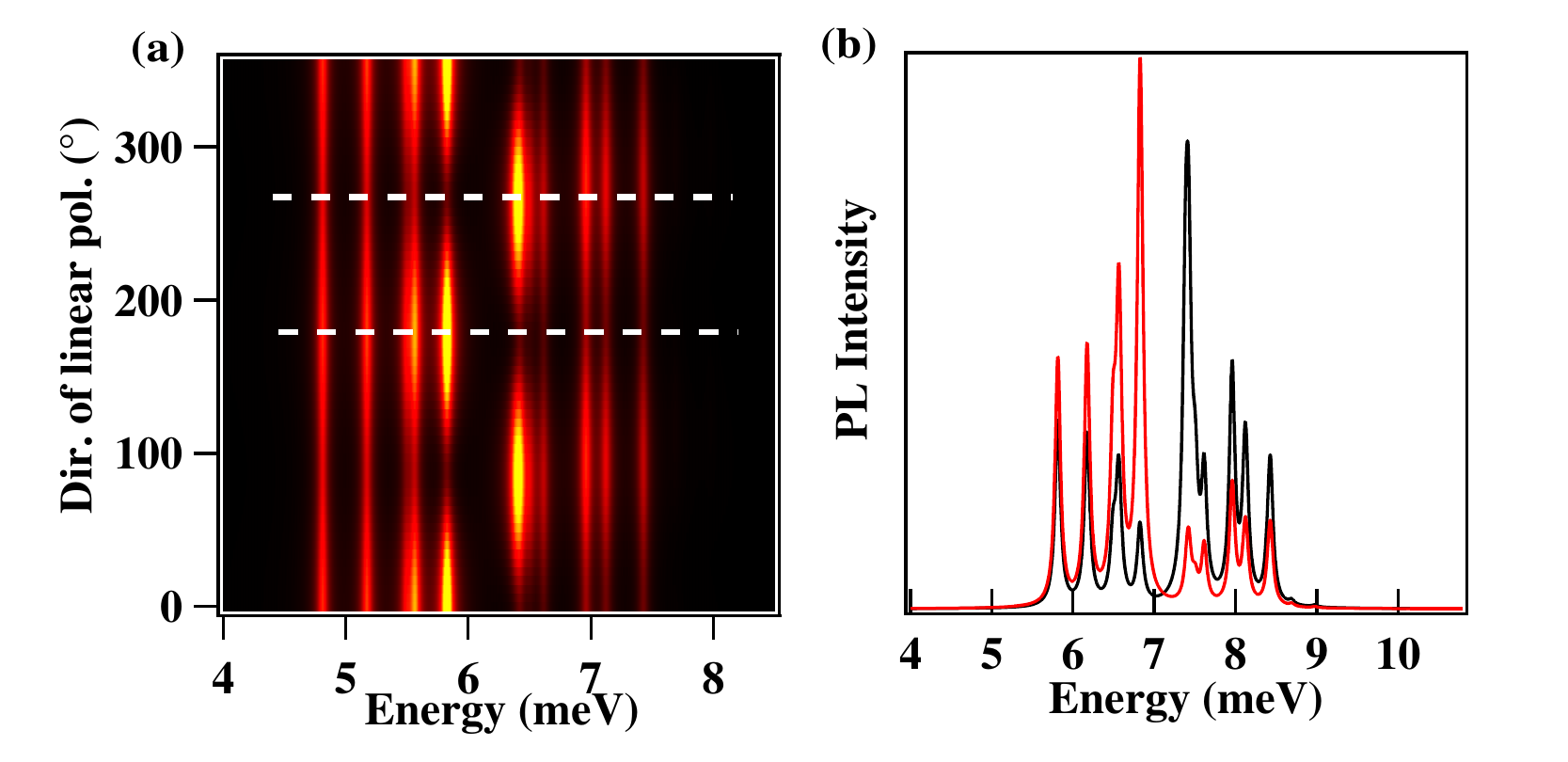}
\caption{Model of the linear polarization properties of XX$^-$-Mn. The direction of linear polarization are indicated with respect with the [100] axis. (a) Intensity map of the linearly polarized PL. (b) PL spectra along two orthogonal directions pointed by dashed lines in (a). Calculation is performed  with $I_{e_1Mn}$= -70$\mu eV$, $I_{e_2Mn}$=0 $\mu eV$, $I_{h_1Mn}$= 230$\mu eV$, $\tilde{\Delta_{0}}$= -700$\mu eV$, $\Delta_{ee}$=-5000$\mu eV$, $\tilde{\Delta_{1}}$= 350$\mu eV$, $\rho_s/\Delta_{lh}$=0.2, $\theta_s$= $\pi/4$, $\xi$=0.15. These values reproduce the order of magnitude of the splittings observed for XX$^-$-Mn in QD2 or QD4.}
\label{FigModXXm}
\end{figure}

The presence of the magnetic atom permits to independently observe two sources of anisotropy: the VBM responsible for the linear polarization in the center of X$^-$-Mn and the long-range exchange interaction in an anisotropic potential responsible for the linear polarization of XX$^-$-Mn. As revealed by the polarization map presented in Fig.~\ref{Fig7bis} for QD2, QD3 and QD4, the directions of linear polarization observed for the two charged complexes are in general not identical. The direction of polarization are measured with respect to the [100] or [010] axis of the sample ({\it i.e.} at 45$^{\circ}$ from the easy cleavage axis of the substrate). In the investigated QDs, the direction of polarization for XX$^-$-Mn are roughly aligned with the [100] or [010] axis. The direction of polarization of X$^-$-Mn are more fluctuating. They can be along the same direction as in QD3, at about 45$^{\circ}$ as in QD4, or at an intermediate angle (around 30$^{\circ}$ in QD2). This directly shows that the shape and strain anisotropy are mainly independent in these self-assembled QDs. 

\begin{figure}[hbt]
\centering
\includegraphics[width=1.0\linewidth]{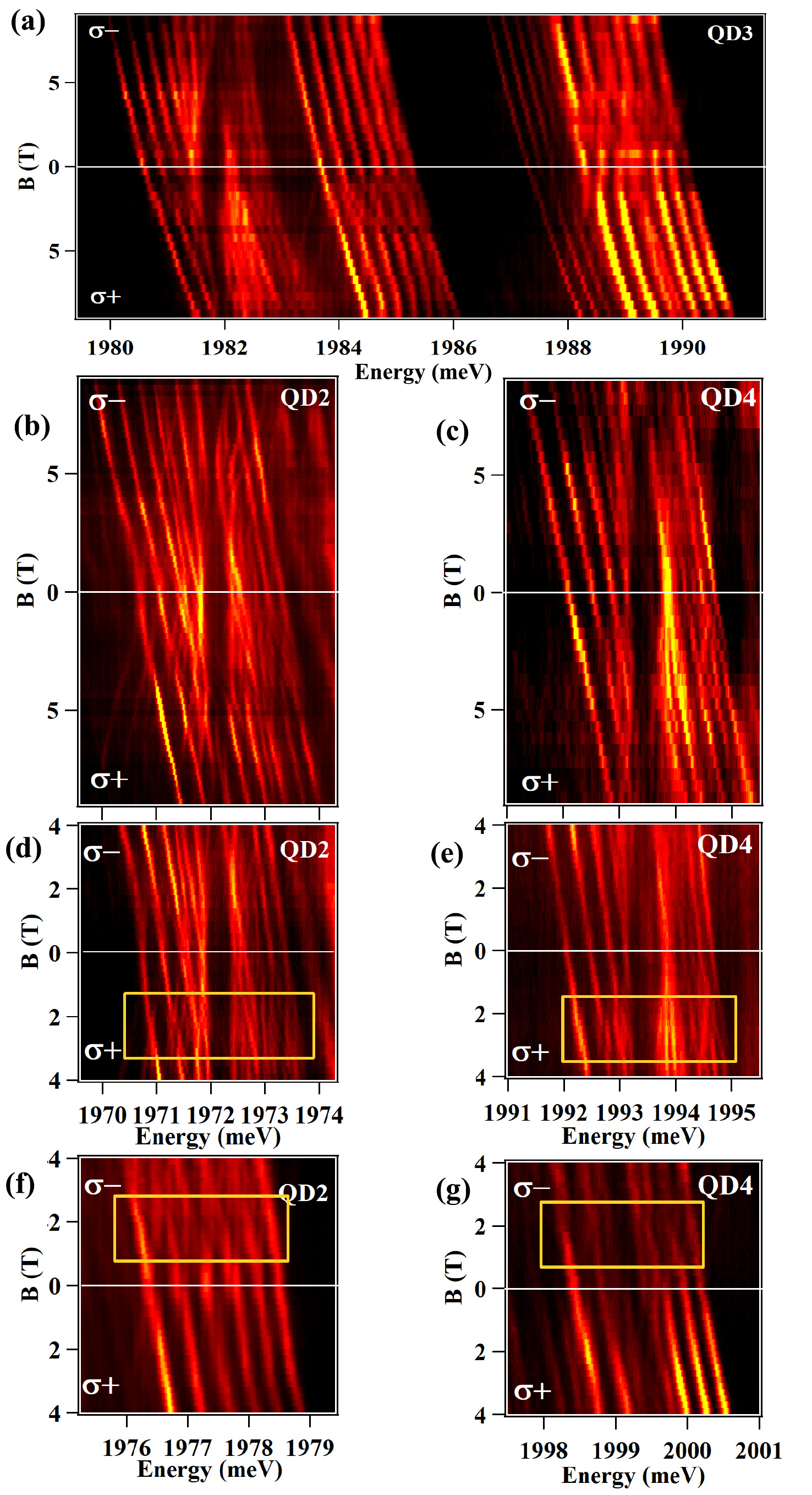}
\caption{PL intensity map of the longitudinal magnetic field dependence of the emission of QD2, QD3 and QD4. (a) Magnetic field dependence of X$^-$-Mn, XX-Mn and XX$^-$-Mn in QD3. (b) and (c): Magnetic field dependence of XX$^-$-Mn in QD2 and QD4 respectively. (d) and (e): Detail of the low field dependence of XX$^-$-Mn in QD2 and QD4. (f) and (g): Detail of the low field dependence of X$^-$-Mn in QD2 and QD4 respectively.}
\label{magneto}
\end{figure}

To analyze in more details the emission XX$^-$-Mn, we use a spin effective model. In the initial state, XX$^-$-Mn can be split by the weak electron-Mn exchange interaction I$_{e_2Mn}\vec{\sigma_2 }\vec{S}$. In the final state, the triplets of X$^{-*}$ are split by the e-h exchange interaction 

\begin{eqnarray}
H_{e_{1}e_{2}h_{1}}=2\tilde{\Delta}_0I_z\sigma^h_z+\tilde{\Delta}_1(I_x\sigma^h_x+I_y\sigma^h_y)
\end{eqnarray}

The exchange constant $\tilde{\Delta_{0,1}}$ differ from the one discussed for the PLE of X$^-$-Mn as the hole is now more strongly localized in the ground state of the dot. One can expect a larger exchange interaction with the electron resulting in a larger splitting of the triplet and a stronger influence of a possible shape anisotropy. The triplet states are further coupled with the Mn atom by the exchange interaction and the X$^{-*}$-Mn Hamiltonian reads:

\begin{eqnarray}
H_{X^{-*}-Mn}=H_{e_{1}e_{2}h_{1}}+ \nonumber \\
I_{e_1Mn}\vec{\sigma_1}\vec{S}+I_{e_{2}Mn}\vec{\sigma_{2}}\vec{S}+I_{h_{1}Mn}\vec{J_{1}}\vec{S}
\label{hamiltXt}
\end{eqnarray}

Under non-resonant optical excitation, the injected e-h pair is not spin polarized and all the XX$^-$-Mn states, weakly split by the e-Mn exchange interaction, are populated with equal probabilities. The emission spectra can be obtained by calculating the overlap of the initial e-Mn states $\langle M,M_z\vert$ with each of the triplet level in the final state of the transition, $\vert T_{\mp1},J_{z_1}=\pm\frac{3}{2},S_z \rangle$, $\vert T_0,J_{z_1}=\pm\frac{3}{2},S_z\rangle$ and $\vert T_{\pm1}, J_{z_1}=\pm\frac{3}{2}, S_z\rangle$. However, transitions towards $\vert T_{\pm1}, J_{z_1}=\pm\frac{3}{2}, S_z\rangle$ are forbidden in a first approximation. For a $\sigma+$ recombination for instance, the intensities of the optically active transitions are given by $\vert\langle M,M_z\vert T_{+1},J_{z_1}=-\frac{3}{2}, S_z\rangle\vert^2$ and $\vert\langle M,M_z\vert T_0, J_{z_1}=-\frac{3}{2}, S_z\rangle\vert^2$. 

Results of a modeling of the PL of XX$^-$-Mn are presented in Fig.~\ref{FigModXXm}. In this model, $\tilde{\Delta_0}$ controls the splitting between the two groups of lines. The long range exchange coupling term $\tilde{\Delta_1}$ controls the central gap and the linear polarization rate. $\tilde{\Delta_1}$ is chosen to be real meaning that the shape anisotropy is oriented along the [100] axis. The main feature of the emission spectra and their linear polarization dependence can be reproduced by the spin effective model. A good agreement can be obtained neglecting the exchange interaction of the magnetic atom with the electron spin in the excited state of the dot. As it will be confirmed by magneto-optic measurements, a VBM induced by in-plane strain anisotropy can produce an additional small gap on the high energy side of the spectra.

\subsection{Magneto-optical properties of the negatively charged biexciton}

The observation of the evolution of the emission of XX$^-$-Mn under magnetic field permits to probe the magneto-optic properties of the charged exciton triplets interacting with the magnetic atom. Magnetic field dependence of different charged excitonic complexes are presented in Fig.~\ref{magneto} for QD2, QD3 and QD4. 

For XX$^-$-Mn, the energy gap induced by $\tilde{\Delta_1}$ which is almost at the center of the emission structure at zero magnetic field evolves towards the low energy side in $\sigma+$ polarization and towards the high energy side in $\sigma-$ polarization. Some anti-crossings are usually observed at weak magnetic field (in the range 2T - 3T) in $\sigma+$ polarization (see detailed weak field scan of QD2 and QD4 in Fig.~\ref{magneto}(d) and (e)). The width of the central gap slightly decreases when a longitudinal magnetic field is applied. This is particularly clear between 0 and 5T.  

\begin{figure}[hbt]
\centering
\includegraphics[width=1.0\linewidth]{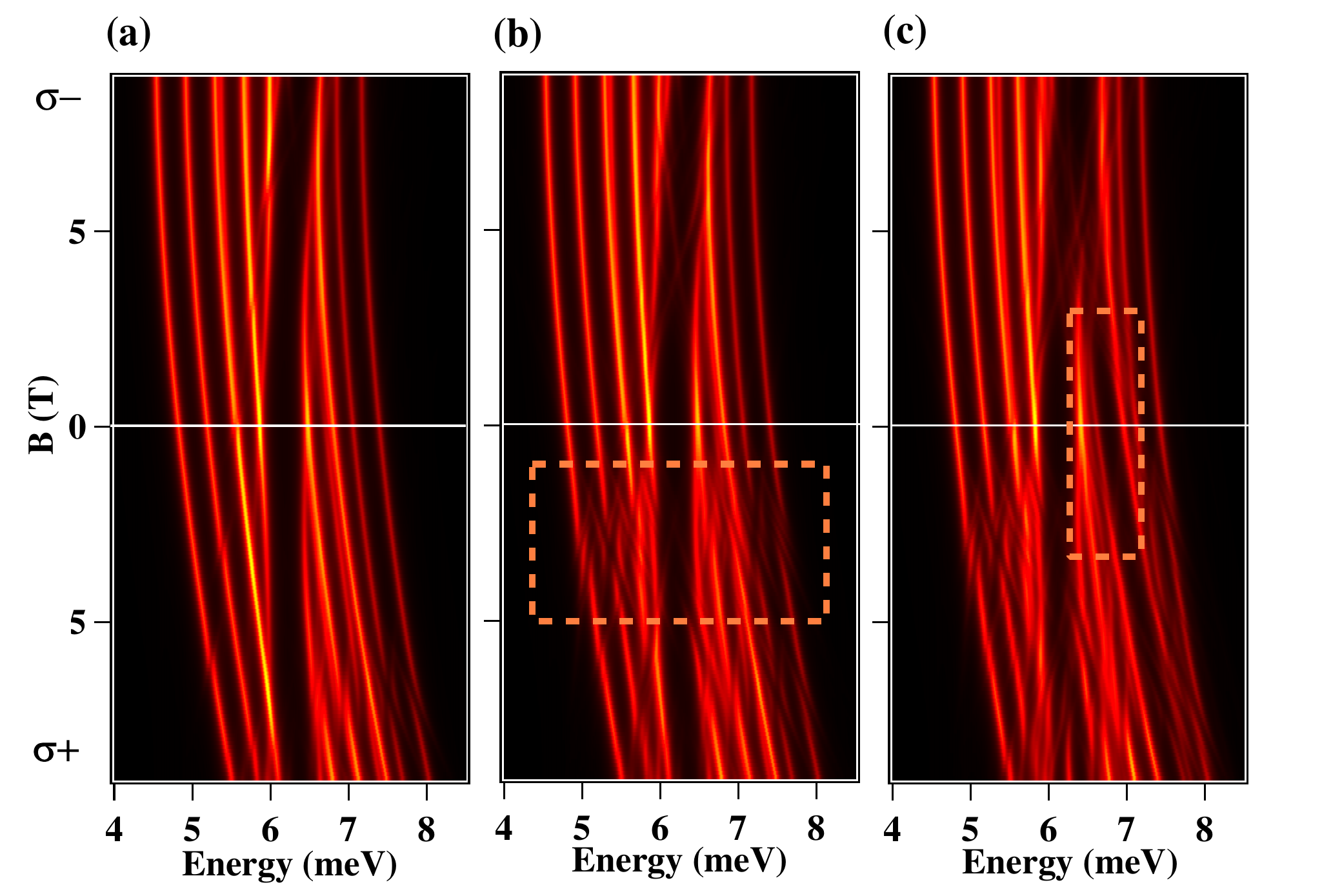}
\caption{(a) Longitudinal magnetic field dependence of the PL intensity of a XX$^-$-Mn calculated with the parameters: (a) $I_{e_1Mn}$=-70$\mu eV$, $I_{e_2Mn}$= 0$\mu eV$,$I_{h_1Mn}$= 230$\mu eV$, $\tilde{\Delta_{0}}$= -700$\mu eV$, $\tilde{\Delta_{1}}$=350 $\mu eV$, $\Delta_{ee}$= -5000$\mu eV$, $\rho_s/\Delta_{lh}$= 0, $\theta_s$= $\pi/4$, $\xi$=0, g$_{e1}$=-0.4, g$_{e2}$=-0.4, g$_h$=0.5, g$_{Mn}$=2, $\gamma$=2.5$\mu eVT^{-2}$. (b) same as (a) with $\xi$=0.15. (d) same as (a) with $\xi$=0.15 and $\rho_s/\Delta_{lh}$= 0.2.}
\label{FigModB}
\end{figure}

The overall behavior of the magnetic field dependence can be reproduced by a spin effective model (Fig.~\ref{FigModB}). For this model, Zeeman energies of the carriers and magnetic atom spins, $g_i\mu_B\vec{S_i}.\vec{B}$, are included in the zero field spin Hamiltonian (\ref{hamiltXt}). A diamagnetic shift of the excitonic complex $\gamma B^2$ is also added. The electron-Mn coupling in the initial state, I$_{e_2Mn}$, is neglected as a value larger than a few tens of $\mu eV$ would produce anti-crossings in the low magnetic field region which are not resolved in the experimental spectra.

The series of anti-crossings observed in the low magnetic field region around 2T-3T in $\sigma+$ polarization can also be reproduced by the model (Fig.~\ref{FigModB}(b)). Similar perturbations are observed for X$^-$-Mn but on $\sigma-$ polarized lines which shift towards low energy under a positive magnetic field (see Fig.~\ref{magneto}(f) and (g)). This is the opposite in the XX$^-$-Mn case.

These levels mixing arise from a VBM induced by shear strain. As in the case of in-plane strain anisotropy, such VBM can be described by effective spin operators acting on the heavy-hole sub-space $\tilde{j}_+=\xi\sqrt{3}\tau_z$, $\tilde{j}_-=\xi^*\sqrt{3}\tau_z$ and $\tilde{j}_z=3/2\tau_z$ \cite{Tiwari2021}. The mixing occurs in the initial state of the optical transition for X$^-$-Mn when the Zeeman energy of the Mn compensate the hole-Mn exchange interaction and hole-Mn levels overlap. The same thing occurs in the final state of the transition in the case of the XX$^-$-Mn when the Zeeman energy of the Mn compensate the bright exciton-Mn exchange splitting (the electron-Mn exchange interaction in the excited state can be neglected).

A perturbation of the spectra in the form of a small gap can appear on the high energy side of the XX$^-$-Mn spectra in some of the dots. As shown by the result of the model presented in Fig.~\ref{FigModB}(c), it is produced by the presence of VBM induced by in-plane strain anisotropy.

\begin{figure}[hbt]
\centering
\includegraphics[width=1.0\linewidth]{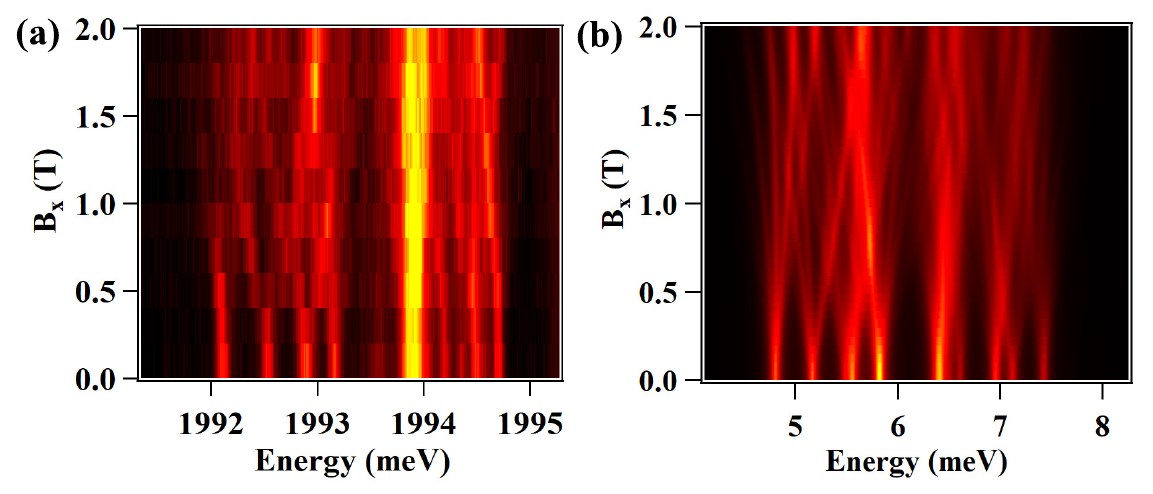}
\caption{(a) Transverse magnetic field dependence of the PL of XX$^-$-Mn in QD4. The magnetic field is applied along the [100] or [010] axis. (b) Model of the transverse magnetic field dependence of XX$^-$-Mn calculated with the same parameters as for Fig~\ref{FigModB}(c) and a magnetic field applied along the [100] axis.}
\label{FigBx}
\end{figure}

The slight decrease of the width of the central gap observed at large magnetic field is not reproduced by the model. It likely corresponds to a decrease of the anisotropic part of the e-h exchange interaction with the increase of a longitudinal magnetic field. A similar reduction of the e-h exchange interaction was already observed in the magneto-PL of a doubly charged exciton coupled with a Mn \cite{Smolenski2015}. This decrease can be induced by a decrease of the anisotropy of the wave-function in the excited state under longitudinal magnetic field which cannot be described by a spin effective model.

As presented in Fig.~\ref{FigBx}(a), under a magnetic field applied in the plane of the QD (transverse magnetic field), each PL line of $XX^-$-Mn splits into multiplets. These splittings result from a mixing of the Mn spin states by the transverse magnetic field. As the Mn levels within XX$^-$-Mn are only weakly coupled with the electron by the exchange interaction, only a weak transverse field split the magnetic atom spin states. They are then quantized in the QD plane along the direction of the applied magnetic field. In the final state and in the range 0T-2T, the Zeeman energy remains weaker than the exchange interaction with the e-h pair confined in the ground state of the QD. The quantization axis remains aligned along the QD growth axis. The change of quantization axis during the optical transition permits any of the initial XX$^-$-Mn level to recombine toward each of the triplet states of the charged exciton. As reproduced by the model presented in Fig.~\ref{FigBx}(b), this significantly increases the number of emission lines.

\section{Conclusion}

To conclude, we examined two ways to probe the exchange interaction of the triplet states of an excited negatively charged exciton with the spin of an individual magnetic atom (Mn) in a QD: the direct excitation of the excited states of X$^-$-Mn and the detection of the PL of XX$^-$-Mn.

The excitation of the triplet state of a charged exciton reveals that the process of negative circular polarization is conserved in the presence of the magnetic atom. This shows that in the investigated QDs, despite the exchange coupling with the magnetic atom, the spin relaxation of the electron triplets remains dominated by the e-h exchange interaction. The emission of XX$^-$-Mn permits to observe independently the isotropic and anisotropic part of the e-h exchange interaction and to analyze the magneto-optic properties of charged exciton triplets coupled with a magnetic atom.

The resonant initialization of excitonic states in the high-energy orbital levels of QDs provides additional degrees of freedom for optical spin-driving protocols. Resonant excitation of charged exciton triplet states in QDs could be used in future work to probe the magnetic field dependence of the spin dynamics of an electron-Mn complex under magnetic field and to control the coupling between two magnetic atoms in a QD interacting with a spin polarized resident electron \cite{Besombes2012,Krebs2013}.

%\begin{acknowledgements}{}

%\end{acknowledgements}

\end{document}